\documentclass[aps,twocolumn,prx,superscriptaddress]{revtex4-1}
\usepackage{graphicx}
\usepackage{dcolumn}
\usepackage{bm}
\usepackage[T1]{fontenc}
\usepackage[latin9]{inputenc}
\usepackage{textcomp}
\usepackage{amsmath}
\usepackage{mathrsfs}
\usepackage{amssymb}
\usepackage{units}
\usepackage{ulem}
\usepackage{color}
\usepackage{soul}

\begin{document}

\title{Exciton-exciton interaction beyond the hydrogenic picture in a MoSe$_2$ monolayer in the strong light-matter coupling regime}

\author{Petr Stepanov}
\affiliation{Univ. Grenoble Alpes, CNRS, Grenoble INP, Institut N\'{e}el, 38000 Grenoble, France}

\author{Amit Vashisht}
\affiliation{Univ. Grenoble Alpes, CNRS, LPMMC, 38000 Grenoble, France}

\author{Martin Klaas}
\affiliation{Technische Physik and Wilhelm Conrad R\"ontgen Research Center for Complex Material Systems, Physikalisches Institut, Universit\"at W\"urzburg, Am Hubland, D-97074 W\"urzburg, Germany}

\author{Nils Lundt}
\affiliation{Technische Physik and Wilhelm Conrad R\"ontgen Research Center for Complex Material Systems, Physikalisches Institut, Universit\"at W\"urzburg, Am Hubland, D-97074 W\"urzburg, Germany}

\author{Sefaattin Tongay}
\affiliation{Arizona State University, Tempe, Arizona 85287 USA}

\author{Mark Blei}
\affiliation{Arizona State University, Tempe, Arizona 85287 USA}

\author{Sven H\"ofling}
\affiliation{Technische Physik and Wilhelm Conrad R\"ontgen Research Center for Complex Material Systems, Physikalisches Institut, Universit\"at W\"urzburg, Am Hubland, D-97074 W\"urzburg, Germany}

\author{Thomas Volz}
\affiliation{Department of Physics and Astronomy, Macquarie University, NSW, 2109, Australia}
\affiliation{ARC Centre of Excellence for Engineered Quantum Systems, Macquarie University, NSW, 2109, Australia}

\author{Anna Minguzzi}
\affiliation{Univ. Grenoble Alpes, CNRS, LPMMC, 38000 Grenoble, France}

\author{Julien Renard}
\affiliation{Univ. Grenoble Alpes, CNRS, Grenoble INP, Institut N\'{e}el, 38000 Grenoble, France}

\author{Christian Schneider}
\affiliation{Technische Physik and Wilhelm Conrad R\"ontgen Research Center for Complex Material Systems, Physikalisches Institut, Universit\"at W\"urzburg, Am Hubland, D-97074 W\"urzburg, Germany}

\author{Maxime Richard}
\affiliation{Univ. Grenoble Alpes, CNRS, Grenoble INP, Institut N\'{e}el, 38000 Grenoble, France}


\begin{abstract}
In transition metal dichalcogenides layers of atomic scale thickness, the electron-hole Coulomb interaction potential is strongly influenced by the sharp discontinuity of the dielectric function across the layer plane. This feature results in peculiar non-hydrogenic excitonic states, in which exciton-mediated optical nonlinearities are predicted to be enhanced as compared to their hydrogenic counterpart. To demonstrate this enhancement, we performed optical transmission spectroscopy of a MoSe$_2$ monolayer placed in the strong coupling regime with the mode of an optical microcavity, and analyzed the results quantitatively with a nonlinear input-output theory. We find an enhancement of both the exciton-exciton interaction and of the excitonic fermionic saturation with respect to realistic values expected in the hydrogenic picture. Such results demonstrate that unconventional excitons in MoSe$_2$ are highly favourable for the implementation of large exciton-mediated optical nonlinearities, potentially working up to room temperature.
\end{abstract}

\maketitle

\begin{figure}[hbt]
\includegraphics[width=0.95\columnwidth]{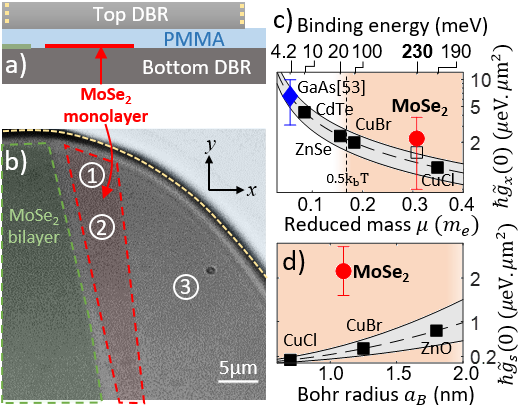}
\caption{\textbf{MoSe$_2$ monolayer microcavity} a) Outline of the microcavity structure: a MoSe$_2$ monolayer (red) and bilayer (green) embedded in PMMA (pale blue) are sandwiched between two Bragg mirrors (DBR). b) white light image of the microcavity; the MoSe$_2$ layers have been color shaded and outlined. The experiments performed at $105\,$K ($127\,K$) were realized in the area labelled 1 (2). The bare cavity measurements have been done in the area labelled 3. c) $\tilde{g}_x(0)$ dependence on $\mu$, and d) $\tilde{g}_s(0)$ dependence on $a_B$. The dashed line and grey area show HE theory corrected by a factor $\alpha_0=3.3\pm 0.8$ following \cite{Estrecho:2019} (see text). The black squares highlight HE theory for CdTe, ZnSe, ZnO, CuBr and CuCl \cite{SI}. The hollow square shows the 30\% enhanced HE theory for MoSe$_2$ \cite{Shahn:2017a}. The blue diamond is a measurement in a GaAs microcavity taken from \cite{Estrecho:2019}. Our best measurements are shown as red circles. The upper axis in c) indicates the bulk exciton binding energy for each materials \cite{SI}. Room temperature-stable excitons are on the right side of the vertical dashed line \cite{SI}.}
\label{fig1}
\end{figure}

The realization of solid-state photonic nanostructures featuring a large third-order optical nonlinearity is a high stake objective. Arrays of coupled nonlinear optical microcavities for instance, would constitute a powerful simulator of nonequilibrium quantum many-body physics \cite{Tomadin:2010a,Noh:2016}, in which phenomena such as a fractional quantum Hall states \cite{Angelakis:2008b,Umcalilar:2012,hafezi:2013}, fermionized states \cite{Carusotto:2009}, nontrivial topological phenomena \cite{Tang:2016}, and a variety of nonequilibrium quantum phase transitions \cite{Tomadin:2010b,Jin:2013,Jin:2014,Raferty:2014} have been predicted. This nonlinearity is also currently a key mechanism in optical communication and computation, as photonic logic gates are mostly built upon it \cite{Andalib:2009,Liu:2011,Mabuchi:2011,Espinosa:2013,Salman:2015}. Increasing further the nonlinearity lowers the energy required to switch the gate, up to a point where the operation works in the quantum regime \cite{Flamini:2019} as required in future quantum computing and communication devices based on photons \cite{Obrien:2009}.

In this context, excitonic states, i.e. bound electron-hole pairs in semiconductor nanostructures, are ideally suited. Their dipole moment provides both strong interaction with light, and a large third order optical nonlinearity due to Coulomb interaction (of magnitude $g_x$) between excitons \cite{Ciuti:1998}, and to the fermionic saturation (of magnitude $g_s$) of the involved electrons and holes \cite{Rochat:2000}.

Photonic waveguides and micropillar microcavities containing quantum dots constitute a successful exploitation of the excitonic nonlinearity \cite{Lodahl:2015a}. The quantum confinement of the excitonic states at the nanometer scale enhances the fermionic saturation, resulting in a giant nonlinearity \cite{Auffeves:2007}. In this system, the conversion of a classical input into various kinds of quantum states of light at cryogenic temperatures has been demonstrated~\cite{Faraon:2008a,Reinhard:2012a,Kim:2013a,Javadi:2015a,Santis:2017a}. However, scaling up such a system is challenging, due to the way semiconductor quantum dots are fabricated \cite{Senellart:2017a}. Another strategy known as polariton quantum blockade \cite{Verger:2006}, gets rid of this difficulty at the expense of a weaker nonlinearity. A semiconductor quantum well (QW) is embedded in the spacer of an optical microcavity such that its excitonic transition is in the strong coupling regime with the cavity resonance \cite{Weisbuch:1992a}. The resulting exciton-polariton states inherit the excitonic nonlinearity, and optimize at the same time the coupling with light \cite{Carusotto:2013}. The onset of the polariton blockade regime has been demonstrated recently in GaAs-based microcavities at cryogenic temperatures ~\cite{Munoz:2019a,Delteil:2019a}.

In order to achieve a robust and practical implementation of exciton-mediated optical nonlinearities, like e.g. for room temperature operation, Arsenide-based semiconductor materials feature a too weak excitonic binding energy $E_b$. $E_b$ is larger in high-bandgap semiconductors, but only at the expense of lower $g_x$ and $g_s$ (see Fig.\ref{fig1}.c-d). This trade-off is governed by the hydrogenic character of the excitonic states in conventional materials, in which $g_x$ and $g_s$ depend only on $E_b$, and $a_B$ the exciton Bohr radius. More explicitly, $g_x=6E_b a_B^2$ \cite{Ciuti:1998}, and $g_s=(4\pi/7)a_B^2\hbar\Omega$ \cite{Rochat:2000}, where $\hbar\Omega$ is the Rabi splitting in the strong coupling regime. In fact, since $E_b a_B^2=\hbar^2/2\mu$, where $\mu=(1/m_e+1/m_h)^{-1}$ is the exciton reduced mass, $g_x$ depends only on the electron and hole effective masses $m_e$ and $m_h$.

In this context, monolayers of semiconductor transition metal dichalchogenides (TMDCs) \cite{Wang:2018a} offer a unique opportunity to break this constrained trade-off. Owing to their atomic-scale thickness, the dielectric constant exhibits a sharp discontinuity across the material plane. The resulting in-plane dependence of the effective Coulomb interaction between electrons and holes is strongly modified, and the resulting excitonic states are non-hydrogenic \cite{Berkelbach:2013,Chernikov:2014a}. Since $g_x$ and $g_s$ are fixed by the spatial characteristic of the Coulomb interaction and of the excitonic wavefunction, we expect both to deviate from the hydrogenic exciton (HE) picture. A 30\% enhancement of $g_x$ is actually predicted in TMDCs as compared to hydrogenic excitons of identical reduced mass \cite{Shahn:2017a}. Signatures of non-negligible excitonic nonlinearities have been observed already in other TMDCs \cite{Barachati_2019}, in charged \cite{Emmanuele_2020} and excited states of excitons \cite{Scuri:2018}, and in polaron-polaritons \cite{Tan:2020}.

In this work, we take advantage of the giant oscillator strength of TMDCs excitons \cite{Poellmann:2015,Robert:2016a,Moody:2015,Korn:2011a} to put a MoSe$_2$ monolayer in the strong coupling regime with the resonance of a microcavity \cite{Dufferwiel:2015a,Sidler:2017a,Lundt:2019a}, and carry out spatially-resolved optical transmission spectroscopy with pulsed laser light, as a function of the intensity. The obtained spectra exhibit signatures of a nonlinear response, from which we derive a quantitative estimate of $g_s$, $g_x$, and the polarization dependence of the latter. The monolithic microcavity that we investigate is shown in Fig.\ref{fig1}.(a,b) with its main features highlighted (See \cite{Lundt:2019a,SI} for details). Its quality factor and Rabi splitting amount to $Q\simeq 730$ and $\hbar\Omega=28\pm3\,$meV respectively. The latter is derived from the anti-crossing of the polariton modes that we observe in a transmission measurement upon sweeping temperature. Details of this characterization can be found in \cite{SI}.



\begin{figure*}[t]
\includegraphics[width=\textwidth]{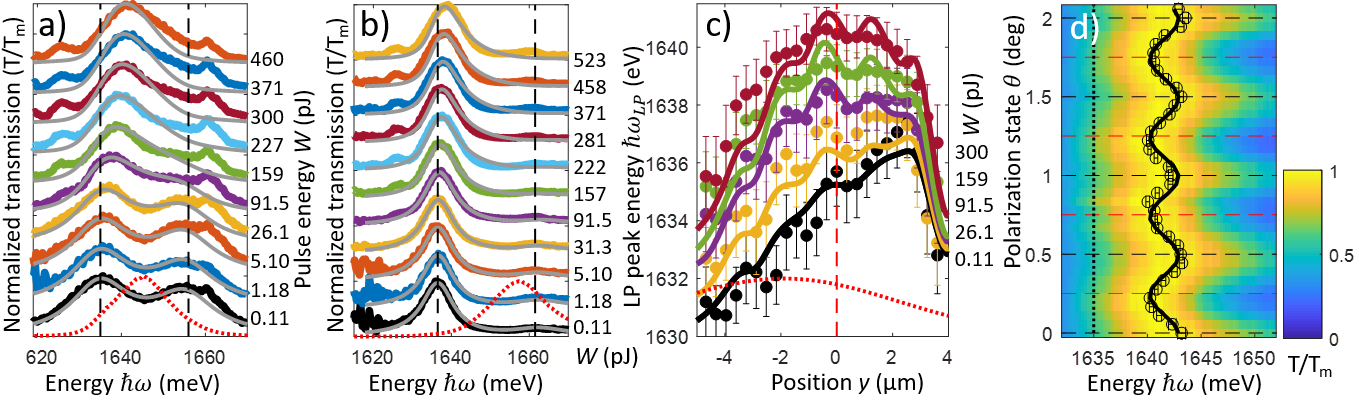}
\caption{\textbf{Nonlinear transmission spectroscopy}. Measured normalized transmission spectra $T(\omega)/{\rm Max}(T)$ at $T=127\,$K (a) and $T=105\,$K (b). The spectra are stacked from the lowest used pulse energy $W$ (bottom) to the highest (top). The pulse energy $W$ used for each spectrum is indicated on the right axis. The laser pulse spectrum is shown in (a,b) as a red dotted line. The dashed vertical black lines in (a,b) highlight the polaritonic resonances in the linear regime. The theoretical fits are shown as solid gray lines. (c) Spatially resolved lower-polariton transmission peak energy measured at $T=127\,$K, across the excitation spot diameter, for increasing $W$ (same color code as in (a)). The spatial laser intensity profile is shown as a red dotted line. The spectra in (a) have been measured at $y=0$ (dashed vertical line). (d) Color-scaled normalized transmission spectra at $W=451\,$pJ versus the polarization state $|\theta\rangle$ (vertical axis). The circle symbols show the lower polariton peak energy; the solid black line is a theoretical fit $B_0+B_\theta\sin ^2 2\theta$, following eqs.(\ref{eq:g_theta},\ref{eq:gs_theta}), where $B_0=7.9\,$meV and $B_\theta=-2.7\,$meV.}
\label{fig2}
\end{figure*}

We then move on to the nonlinear transmission measurements. We use a pulsed Ti:sapphire laser that delivers $\sim 200\,$fs pulses with a spectrum $I_{\rm las}(\omega)$ of tunable mean energy $\hbar\omega_{\rm las}=[1640,1660]\,$meV and a bandwidth $\gamma_{\rm las}\sim 10\,$meV. Its purpose is the ultrafast creation of a dense polariton population resonantly, without overheating the sample, and to perform a broadband transmission spectrum measurement which is able to capture both the lower and upper polaritons. Heating from residual absorption is further suppressed by chopping the laser into a $0.7\%$ on-off duty-cycle. The beam is prepared into a Gaussian mode which is focused on the microcavity surface into a $\sigma=5.8\,\mu$m waist size spot. We use a quarter-wave plate to tune the laser polarization among the states $|\theta\rangle$, where $|\theta\rangle=\sin(2\theta)/\sqrt{2}|x\rangle+\left[i+\cos(2\theta)\right]/\sqrt{2}|y\rangle$, $|x,y\rangle$ is the linear polarization basis oriented as shown in Fig.\ref{fig1}.(b), and $\theta$ is the wave plate rotation angle with respect to $y$. In the first part of this work we use $y$-polarized light ($\theta=0$). The time-integrated transmitted light intensity $I_T$ is collected with a microscope objective, and imaged at the entrance focal plane of a 300 grooves/mm grating spectrometer. By doing so, we obtain space and frequency-resolved transmission spectra $T(\omega,y)=I_T(\omega,y)/I_{\rm las}(\omega,y)$. Since our aim is to provide a quantitative estimate of the interactions, we also need to know the electromagnetic energy $W$ in each pulse. To do so, the time-averaged laser power $P_{\rm las}$ is measured at the input cryostat window, just before the laser light impinges the cavity backside using a thermal-head powermeter. $T(\omega,y)$ is thus measured from $W_0=0.1\,$pJ, which is well below the onset of the nonlinear regime, up to several hundreds of pJ, which is well above.

Two such measurements, realized in area (1) and (2), at temperatures $T=127\,$K and $T=105\,$K, respectively, are shown in Fig.\ref{fig2}.a and Fig.\ref{fig2}.b. The excitonic fraction of the polariton field in each case is $|X|^2=0.48$ and $|X|^2=0.33$, respectively. We indeed exploit the fact that the excitonic transition energy is temperature-dependent to control the detuning $\Delta(T)=\omega_{c,0}-\omega_x(T)$ between the bare cavity (frequency $\omega_{c,0}$) and the excitonic level (frequency $\omega_x$) \cite{SI}, and hence the excitonic fraction $|X|^2$ ($|C|^2=1-|X|^2$) of the lower (upper) polariton states \cite{Carusotto:2013}. The effective polariton-polariton interaction constant is thus varied as it depends on $|X|^2$ \cite{Rochat:2000,SI}. The laser pulse spectral overlap with the polariton modes is also different in the two experiments. We take advantage of these variations to test the robustness of our quantitative estimate of $g_s$ and $g_x$, as they should not depend on these parameters. The plotted transmission spectra are normalized to their maximum $T_m$ for clearer representation. In the linear regime (bottom spectra, $W=0.11\,$pJ), we observe both the upper and lower polariton resonances, with a mostly equal weight at $T=127\,$K, and with a dominant lower polariton peak at $T=105\,K$, consistently with their respective photonic fraction. Two smaller peaks are also visible in these spectra at $\hbar\omega=1625.6\,$meV and $\hbar\omega=1660.4\,$meV that we traced back, by real space analysis, to bare cavity resonances situated within the small gap separating the MoSe$_2$ monolayer from the bilayer. Upon increasing $W$, the polaritonic resonances exhibit a clear and consistent trend: at moderate $W$, the lower polariton peak blueshifts, while the upper polariton essentially does not. This behaviour difference is key to distinguish between the contributions of $g_s$ and $g_x$ to the nonlinearity. Indeed, while Coulomb interaction contributes to blueshift both lower and upper polaritons, the saturation causes a reduction of the effective Rabi spitting, and thus shifts the lower and upper polaritons in opposite directions \cite{SI}. The trend we observe thus indicates that the saturation contributes significantly to the nonlinearity, consistently with recent reports \cite{Gu:2020}.

We interpret these spectra quantitatively, by theoretical simulation of the polariton field ultrafast evolution, including the shape of the laser pulse in time and space. Specifically, we derive a mean-field input/output theory in the exciton-photon basis \cite{SI}, including exciton-exciton interactions and saturation effects. Owing to the exciton state properties, $g_x$ has two contributions: $g_{x,\parallel}$ and $g_{x,\perp}$ corresponding to the interactions between parallel and opposite spin excitons, that couple to co- and cross-circularly polarized light. After transformation into the $\theta$ polarization basis, the equation of motion for the exciton and photon fields $\psi_{\rm x,c}(\mathbf{r},t)$ read:
\begin{eqnarray}
\label{eq:mf1}
i\partial_t \psi_{\rm c} &=& \left(\omega_{c,0}-\frac{\hbar}{2m}\nabla^2 - i\frac{\gamma_c}{2}+V_c(x,y)\right)\psi_{\rm c}  \nonumber \\
&& + \left(\frac{\Omega}{2}-\frac{\tilde{g}_s(\theta)}{2}|\psi_{\rm x}|^2 \right)\psi_{\rm x}+\sqrt{2\gamma_{\rm in}}A_{\rm in} \\
\label{eq:mf2}
i\partial_t \psi_{\rm x} &=& \left(\omega_x - i\frac{\gamma_x}{2}+\tilde{g}_x(\theta)|\psi_{\rm x}|^2\right)\psi_{\rm x}  \nonumber \\
&& + \left(\frac{\Omega}{2}-\tilde{g}_s(\theta)|\psi_{\rm x}|^2 \right)\psi_{\rm c}-\frac{\tilde{g}_s(\theta)}{2}\psi_x^2\psi_c^*,
\end{eqnarray}
where $A_{\rm in}(\mathbf{r},t)$ is the $\theta$-polarized incident laser pulse field density, $\omega_{c,0}/2\pi$ is the bare cavity resonance frequency at vanishing in-plane wavevector $k_\parallel=0$, $m$ its effective mass, and $V_c$ is the potential describing the spatial dependence $\omega_{c,0}$, $\omega_x/2\pi$ is the excitonic transition frequency, of which we can neglect the kinetic contribution and $\hbar\Omega$ is the Rabi splitting, $\gamma_x$ is the excitonic non-radiative relaxation rate, $\gamma_c=\gamma_{\rm in}+\gamma_{\rm out}$ is the cavity radiative decay rate, and $\gamma_{\rm in}$ ($\gamma_{\rm out}$) are the cavity coupling rate on the laser input side (of the transmission side). Finally, $\tilde{g}_s(\theta)$, and $\tilde{g}_x(\theta)$ are the saturation and exciton-exciton interaction constants in the $\theta$-polarization basis, given by \cite{SI}
\begin{eqnarray}
\label{eq:g_theta}
&\tilde{g}_x(\theta)&=\frac 1 2 \left(g_{x,\parallel}+g_{x,\perp}\right)+\frac 1 2 \left(g_{x,\parallel}-g_{x,\perp}\right) \sin^2(2\theta)\\
\label{eq:gs_theta}
&\tilde{g}_s(\theta)&=\frac{g_s}{2}(1+\sin^2(2\theta))
\end{eqnarray}
Note that in Eq.(\ref{eq:mf1}-\ref{eq:mf2}) we have neglected the contribution of the cross-polarized $\bar \theta$ components of the exciton and photon fields since the laser excites only one $\theta$ component, and the interactions terms provide only density-mediated couplings which vanish if one of the two fields is zero. We also checked experimentally that the polariton modes exhibit no birefringence \cite{SI}.

In order to fully account for the time profile of the excitation pulse, and of the Gaussian shape of the spot in real space, we solved this model numerically. The experimental parameters entering the model are the microcavity and laser characteristics, which are known accurately. The interaction constants $\tilde{g}_x(\theta)$ and $\tilde{g}_s(\theta)$, are thus the only free parameters. We first apply this model to the spectra shown in Fig.\ref{fig2}.(a,b) (in the $\theta=0$ polarization state). $\tilde{g}_x(0)=\left(g_{x,\parallel}+g_{x,\perp}\right)/2$ and $\tilde{g}_s(0)=g_s/2$ are thus derived with their uncertainty by numerical optimization of the fit between the model and the measurements \cite{SI}. This analysis yield $\tilde{g}_x(0)=2.2\pm 1.6\,\mu$eV.$\mu$m$^2$ and $\tilde{g}_s(0)=2.16\pm 0.5\,\mu$eV.$\mu$m$^2$ for the experiment at $T=127\,$K shown in Fig.\ref{fig2}.a. The experiment at $T=105\,$K consistently yields $\tilde{g}_x(0)=4.3\,\mu$eV.$\mu$m$^2$, and $\tilde{g}_s(0)=1.6\,\mu$eV.$\mu$m$^2$, albeit with a much larger uncertainty due to the fact that the upper polariton contribution to the spectra is small, and hence prevents determining accurately the relative contribution of $\tilde{g}_x(0)$ and $\tilde{g}_s(0)$. We also derive the excitonic densities (half-width-at-half-maximum in time and space) that increases from $5\times 10^8\,$cm$^{-2}$ ($W=0.11\,$pJ) to $9\times 10^{11}\,$cm$^{-2}$ ($W=460\,$pJ). Note that at high $W$, the saturation effect is large and our model is expected to overestimate it in this regime \cite{Rochat:2000,Combescot:2008,Kyriienko_2019,Emmanuele_2020}. This is indeed the trend that we observe in the last four spectra in Fig.\ref{fig2}.b, in which the theory predicts a slightly smaller Rabi splitting than in the experiment. Yet, except for this feature, the spectral shape and peak energies evolution for increasing $W$ are in very good agreement with the experiment.

We cross-checked this quantitative analysis by looking at another footprint of the nonlinearity: the nontrivial spatially-dependent transmission spectrum $T(y,\omega)/T_m$ that results from the interplay between the Gaussian shape of the spot and the nonlinearity. Fig.\ref{fig2}.c shows the lower polariton transmission peak energy $E_{\rm lp}(y)$, plotted versus $y$, where $y$ is the position along a diameter of the laser spot, and $y=0$ is the laser spot intensity maximum position. The lowest spectrum (black) is obtained in the linear regime ($W=0.11\,$pJ) and thus shows the lower polariton potential $V(y)$, from which we derive $V_c(y)$. For increasing $W$ the nonlinearity changes this shape as the blueshift depends on the local density and excitonic fraction. We can fit this behaviour quantitatively with our model, and a good agreement is obtained for $\tilde{g}_x(0)= 4.3^{+30}_{-4}\,\mu$eV.$\mu$m$^2$, and $\tilde{g_s}(0)= 3.2\pm 0.8,\mu$eV.$\mu$m$^2$. The large uncertainty reflects the fact that the upper polariton contribution is weak in the dataset, and the relative contributions of ${g}_x(0)$ and ${g}_s(0)$ are hard to distinguish. Yet, the result is consistent with the spectral analysis.

We verified that the nonlinearities that we measure in this work come from the monolayer and not from any other materials within the structure. We thus measured $T(\omega)/T_m$ in area 3, which is a bare cavity free from MoSe$_2$. The area exhibits a sharp cavity mode, that does not shift ($\hbar\delta\omega_{\rm c}(W)=0\pm 0.025\,$meV) up to the highest applied pulse energy ($W=1.12\,$nJ), as is shown in detail in \cite{SI}.

In Fig.\ref{fig1}.c and Fig.\ref{fig1}.d, we plotted the theoretical HE interaction constants $\tilde{g}_x(0)\simeq 3\alpha_0\hbar^2/2\mu$ (in which we assumed that $|g_\perp|\ll g_\parallel$) versus $\mu$, and $\tilde{g}_s(0)=\alpha_0^2(2\pi/7)\hbar\Omega a_B^2$ versus $a_B$ (dashed lines). $\alpha_0=3.3\pm0.8$ is introduced in order for the theory to agree quantitatively with the measurement in Estrecho et al. \cite{Estrecho:2019}, where $g_{x,\parallel}=13\pm 3.4\,\mu$eV.$\mu$m$^2$ is found for a planar microcavity with GaAs quantum wells \cite{SI}. This deviation might arise from the strict 2D approximation of the excitonic wavefunction in the theory, which is likely inaccurate in realistic quantum wells \cite{Estrecho:2019}. Using excitonic reduced masses from the literature \cite{SI}, a few materials are highlighted (squares) along these theoretical curve. In Fig.\ref{fig1}.c, the bulk exciton binding energies are also indicated for each material on the top axis as reference \cite{SI}. The measurements obtained from the analysis of Fig.\ref{fig2}.a are shown as a red circle in Fig.\ref{fig1}.c-d. Our measured $\tilde{g}_x(0)$ is found to moderately exceed HE's theory, and is fully compatible with the $30\%$ enhancement (hollow square in Fig.\ref{fig1}.c) predicted in \cite{Shahn:2017a}, while $\tilde{g}_s(0)$ exceeds HE's theory by a large factor $7\pm2$. A possible origin of this larger deviation is already visible in the HE picture, in which $g_s$ depends directly on the dielectric function square (via $a_B$), while $g_x$ essentially does not.

We finally characterized the spin anisotropy of the nonlinearity at $T=127\,$K, during the same experimental run as that shown in Fig.\ref{fig2}.a, by measuring the transmission spectrum versus $\theta$. The results are shown in Fig.\ref{fig2}.d: upon increasing $\theta$ from $0$ (linear polarization) to $\pi/4$ (circular polarization) at a fixed $W=451\,$pJ, the spectrum exhibits a global redhsift of $2.7\,$meV. Using our model and Eqs.(\ref{eq:g_theta},\ref{eq:gs_theta}) \cite{SI}, this behaviour implies that  $g_{x,\perp}$ is about twice larger than $g_{x,\parallel}$, and positive. In TMDC monolayers \cite{Shahn:2017a}, like in conventional materials, the Coulomb interaction between polaritons is in principle dominated by exchange interaction \cite{Ciuti:1998}, for which $g_{x,\perp}$ is expected to be negative and small as compared to $g_{x,\parallel}$ \cite{Glazov:2009,Vladimirova:2010}. Our result differs from
this picture, and is thus highly non-trivial. Its precise interpretation requires a fully dedicated investigation that exceeds the scope of the present work.

A possible explanation could be the involvement of an intermediate state, like spin-2 dark excitons \cite{Glazov:2009}, or biexcitons \cite{Carusotto:2010a,Takemura:2017}. In such a mechanism, $g_{x,\perp}$ is enhanced and takes a positive sign when the two-polaritons state is close, and on the high energy side of the intermediate state. In a MoSe$_2$ monolayer, the dark exciton state is a few meV above the bright one \cite{Robert:2020}, such that the upper polariton state, nominally $12.2\,$meV above the bright exciton, could benefit from this resonance at the peak intensity, when the saturation brings it closer. A resonance with the biexciton state is expected $10\,$meV \cite{Hao:2017,Bleu:2020} below the bright exciton, which is $3\,$meV above the nominal energy of the lower polariton, and thus also favourable at the peak intensity. Finally, at such large $W$, higher order many-body correlations and the composite nature of excitons might start to contribute, such that our estimate of $g_{x,\perp}$ might be too inaccurate. Yet, owing to the robustness that the model has demonstrated in reasonably capturing the measurements in Fig.\ref{fig2}.(a-c), we expect that this estimate is at least qualitatively correct; namely, that $g_{x,\perp}$ is positive and comparable in magnitude to $g_{x,\parallel}$ and $g_s$.

In summary, we have shown that a MoSe$_2$ monolayer in the strong coupling regime displays enhanced exciton-mediated optical nonlinearity as compared to comparable HE excitons, in particular via the excitonic saturation mechanism. We also observe a non-trivial spin anisotropy of the interaction which deserves future investigation. Our results demonstrate that non-hydrogenic exciton in MoSe$_2$, and potentially in other TMDC materials, offer new perspectives for the engineering of exciton-mediated optical nonlinearities.

\begin{acknowledgments}
PS and AV contributed equally to this work. The authors acknowledge fruitful discussions with D. Basko, O. Kyriienko and D. Ferrand. P.S., M.R. and J.R. are supported by the French National Research Agency in the framework of the Investissements d'Avenir program (ANR-15-IDEX-02) and by the research grant ANR-16-CE30-0021. S.T. acknowledges support from NSF DMR 1838443 and ARO Materials STIR program. M.K., N.L., S.H., and C.S. acknowledges support by the State of Bavaria. C.S. acknowledges support by the European Research Commission (ERC, Project unLiMIt-2D, 697228). T.V. acknowledges the ARC Centre of Excellence for Engineered Quantum Systems (CE170100009). A.V. acknowledges the European Union Horizon 2020 research and innovation programme under the Marie Sk\l{}odowska-Curie grant agreement No 754303.
\end{acknowledgments}

\end{document}


\title{SUPPLEMENTAL MATERIAL - Exciton-exciton interaction beyond the hydrogenic picture in a MoSe$_2$ monolayer in the strong light-mater coupling regime}

\author{Petr Stepanov}
\affiliation{Univ. Grenoble Alpes, CNRS, Grenoble INP, Institut N\'{e}el, 38000 Grenoble, France}

\author{Amit Vashisht}
\affiliation{Univ. Grenoble Alpes, CNRS, Grenoble INP, LPMMC, 38000 Grenoble, France}

\author{Martin Klaas}
\affiliation{Technische Physik and Wilhelm Conrad R\"ontgen Research Center for Complex Material Systems, Physikalisches Institut, Universit\"at W\"urzburg, Am Hubland, D-97074 W\"urzburg, Germany}

\author{Nils Lundt}
\affiliation{Technische Physik and Wilhelm Conrad R\"ontgen Research Center for Complex Material Systems, Physikalisches Institut, Universit\"at W\"urzburg, Am Hubland, D-97074 W\"urzburg, Germany}

\author{Sefaattin Tongay}
\affiliation{Arizona State University, Tempe, Arizona 85287 USA}

\author{Mark Blei}
\affiliation{Arizona State University, Tempe, Arizona 85287 USA}

\author{Sven H\"ofling}
\affiliation{Technische Physik and Wilhelm Conrad R\"ontgen Research Center for Complex Material Systems, Physikalisches Institut, Universit\"at W\"urzburg, Am Hubland, D-97074 W\"urzburg, Germany}

\author{Thomas Volz}
\affiliation{Department of Physics and Astronomy, Macquarie University, NSW, 2109, Australia}
\affiliation{ARC Centre of Excellence for Engineered Quantum Systems, Macquarie University, NSW, 2109, Australia}

\author{Anna Minguzzi}
\affiliation{Univ. Grenoble Alpes, CNRS, Grenoble INP, LPMMC, 38000 Grenoble, France}

\author{Julien Renard}
\affiliation{Univ. Grenoble Alpes, CNRS, Grenoble INP, Institut N\'{e}el, 38000 Grenoble, France}

\author{Christian Schneider}
\affiliation{Technische Physik and Wilhelm Conrad R\"ontgen Research Center for Complex Material Systems, Physikalisches Institut, Universit\"at W\"urzburg, Am Hubland, D-97074 W\"urzburg, Germany}

\author{Maxime Richard}
\affiliation{Univ. Grenoble Alpes, CNRS, Grenoble INP, Institut N\'{e}el, 38000 Grenoble, France}

\date{\today}

\maketitle

\section{Experimental setup}

\begin{figure}[!b]
\includegraphics[width=\columnwidth]{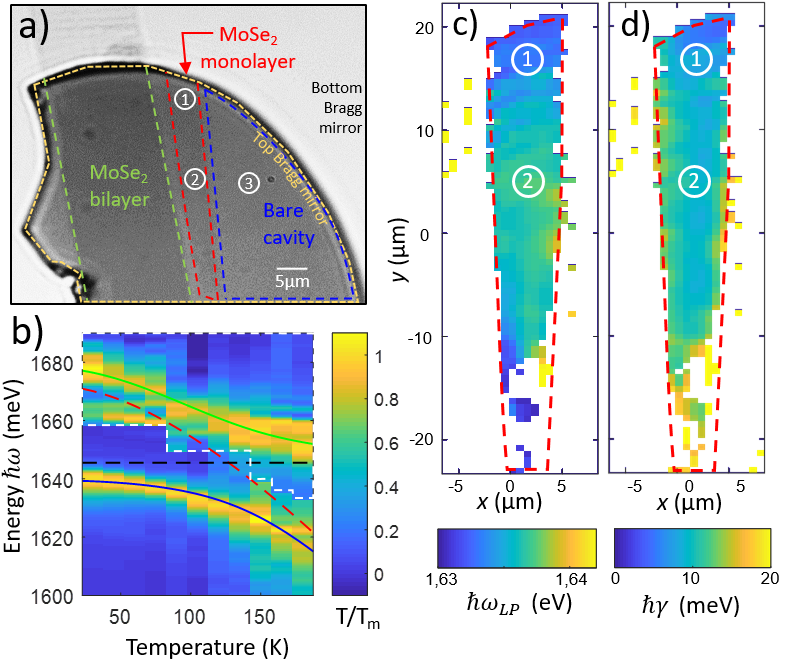}
\caption{\textbf{Microcavity characterization}. a) White light transmission image of the microcavity. The main cavity structures are highlighted with colored lines. White: top mirror, dashed green: MoSe$_2$ bilayer, dashed red: MoSe$_2$ monolayer, dashed blue: bare cavity structure (PMMA spacer). The circled numbers are the regions of interest labels (cf. main text). b) White light transmission spectra (vertical) plotted in color scale versus temperature (horizontal). The transmission in the area above the white dashed line has been enhanced by numerical factors increasing from 1.5 (T=128K) to 30 (T=30K) for decreasing temperatures. The upper and lower polariton modes, the bare excitonic transition, and the bare cavity mode calculated in a coupled oscillators model, are shown in solid green and black, dashed red, and dashed black respectively. (c,d) Real-space color-scaled maps of the lower polariton transmission peak central energy $E_{\rm lp}(x,y)$) (c), and linewidth $\hbar\gamma(x,y)$ (d).}
\label{figS2}
\end{figure}

\begin{figure*}[t]
\includegraphics[width=0.8\textwidth]{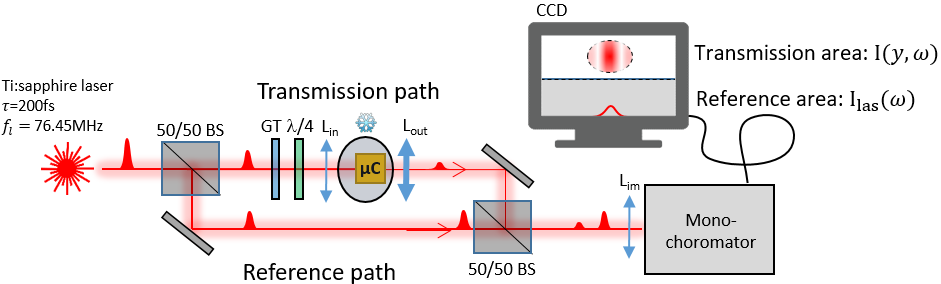}
\caption{\textbf{Experimental setup} The leftmost symbol represents the pulsed Ti:sapphire laser. "BS": Beamsplitter; "GT": Glan Thomson linear polarizer used to clean up the laser polarization. "$\lambda/4$": quarter-wave plate. It can be rotated by an angle $\theta$ for the laser beam polarization preparation into the states $|\theta\rangle$. $L_{in}$: 5-cm doublet lens for the laser beam focusing on the microcavity surface. "$\mu$C" is the microcavity, fitted in a variable temperature cryostat. $L_{out}$ is the collection and imaging objective of $f=4.5\,$mm. "$L_{im}$" is a $250\,$mm focal length doublet lens used to realize the transmission image in the entrance focal plane of the monochromator. "CCD" is the charge-coupled device camera that collects the monochromator dispersed image.}
\label{figS1}
\end{figure*}

\subsection{Microcavity structure and characterization}
The microcavity is obtained by transferring an exfoliated MoSe$_2$ monolayer by dry-gel method \cite{Castellanos:2014a} onto a SiO2/TiO2 bottom DBR (10 pairs, stop band center at 750 nm). We then spin-coated a 126 nm thick poly-methyl-methacrylate (PMMA) buffer layer, and placed a piece of a separate SiO2/TiO2 DBR with lateral dimensions of a few tens of microns (8.5 pairs, stop band center at 750 nm) in order to close the cavity. The Transmission spectra show that a quality factor of $Q\simeq 730$ is thus achieved, which correspond to a bare cavity photon decay rate $\gamma_c=2.2\,$meV/$\hbar$. This decay is split into two imbalanced channels $\gamma_{in}$ and $\gamma_{out}$, corresponding respectively to back mirror coupling (laser input side), and front mirror coupling (emission collection side). A transfer matrix calculation shows that $\gamma_{out}/\gamma_{in}\simeq 3$.

The upper and lower polariton modes of the microcavity are characterized in Reflectivity for different temperature from $T=10\,$K to $T=200\,$K, at position ROI (1) (cf. Fig.1.b in the main text). The temperature dependence of the excitonic transition energy is well captured by the phenomenological equation \cite{Arora:2015a}
\begin{equation}
    \hbar\omega_x(T)=\hbar\omega_{x,0}-\frac{7.5\times 10^{-4}T^2}{T+300},
\end{equation}
The temperature dependence of the cavity mode $\hbar\omega_c$ is assumed to be comparatively small. The temperature-dependent upper and lower polariton modes are simulated with a the two-coupled oscillator model
\begin{equation}
    \omega_{UP,LP}(T)=\frac{\omega_c+\omega_x(T)}{2}\pm\frac{1}{2}\sqrt{\left[(\omega_c-\omega_x(T)\right]^2+\Omega^2},
\end{equation}
where $\hbar\Omega$ is the Rabi splitting. By fitting the reflectivity in Fig.\ref{figS2}.b, we obtain $\hbar\omega_{X,0}=1672\pm1.2\,$meV, $\hbar\omega_c=1645.5\pm1.8\,$meV at this particular point, and $\hbar\Omega=28\pm3\,$meV.

We also characterize the homogeneity of the MoSe$_2$ monolayer by spatially-resolved transmission. To do so, we fixed the temperature at $T=130\,$K and sweep a CW laser across the lower polariton transmission peak and record images of the transmitted light $T(x,y,\omega_{\rm las})$ for every laser frequencies. The thus obtained spectra at each position $(x,y)$ are then fitted with a Lorentzian lineshape to determine the lower polariton peak energy $E_{LP}$, and full-width-at-half-maximum (FWHM) $\hbar\gamma$. The resulting maps are shown in Fig.\ref{figS2}.c and Fig.\ref{figS2}.d respectively. We see that the lower polariton resonance energy is subject to a moderate energy gradient, with a $\sim 10\,$meV energy increase from the monolayer top to bottom, and a $10\,$meV average full-width-at-half-maximum (FWHM) $\gamma$, which is rather narrower in the central part of the monolayer. The maps also show that the lower tip of the monolayer (the area below $y=-10\,\mu$m) is damaged, and show no polaritonic resonance.

\subsection{Experimental method}
The principle of the micro-transmission setup is shown in Fig.\ref{figS1}. It consists in a Mach-Zender interferometer, where the microcavity, cooled down in a variable temperature cryostat, occupies one of the arm (the "transmission arm" in Fig.\ref{figS1}). The other arm ("reference arm" in Fig.\ref{figS1}) provides a reference spectrum of the laser pulse. In the transmission arm, the laser beam is focused into a $\sim 5\,\mu$m waist gaussian spot with a 5cm focal doublet. The resulting transmission image is collected with a 40x microscope objective of $4.5\,$mm focal length. The two beams are collected with a 50/50 beam splitter, and sent into a 75cm focal monochromator with a 300 grooves/mm grating fitted with a CCD camera, on a slightly different path, such that each beam generates a separated image on the CCD camera. The lower CCD area shows the reference beam (reference path), that provides the laser spectrum $I_{las}(\omega)$ and intensity (relative to the the transmission path).The upper area, shows a spectrally resolved vertical cross-section of the transmission image $I(\omega,y)$.


\subsection{Fitting procedure and numerical determination of $\tilde{g}_x(0)$ and $\tilde{g}_s(0)$ and their uncertainty}

In order to determine the best fit of the model with the data, we use an optimization algorithm that minimizes the residues $R^2$ between the experimental data and the model, where
\begin{equation}
R^2(\tilde{g}_x,\tilde{g}_s)=\frac{1}{N-2}\displaystyle\sum_{j=1}^{N} \left[T_{\rm exp,j}-T_{\rm th}(\tilde{g}_x,\tilde{g}_s,\omega_j)\right]^2,
\end{equation}
$T_{\rm exp,j}$ is the measured transmission at the frequency $\omega_j/2\pi$, and $T_{\rm exp,j}$ is the calculated transmission at this frequency. The function $R^2(\tilde{g}_x,\tilde{g}_s)$ is systematically swept within a reasonable range, and the best fit $(\tilde{g}_{x,\rm fit},\tilde{g}_{s,\rm fit})$ is determined as the point for which $R^2(\tilde{g}_x,\tilde{g}_s)$ reaches its minimum $R^2_{\rm min}$. The $1\sigma$ uncertainty on the best fit is estimated as
\begin{equation}
\delta \tilde{g}^2=\frac{(N-2)R^2_{\rm min}}{\displaystyle\sum_{j=1}^{N}\left[\frac{\partial}{\partial \tilde{g}}T_{\rm th}(\tilde{g}_{x,\rm fit},\tilde{g}_{s,\rm fit},\omega_j)\right]^2}.
\end{equation}
For the spectral fits (Fig.2.a-b of the main text), the fitted observable is the transmission $T$. For the spatial fit (Fig.2.c of the main text), the fitted observable is the lower polariton peak maximum $\hbar\omega_{LP}$. The latter is obtained from the experimental data using an order-5 polynomial fit of the spectrum region close to the lower polariton transmission peak maximum.

\section{Polarization-resolved measurements}

\begin{figure*}[t]
\includegraphics[width=0.8\textwidth]{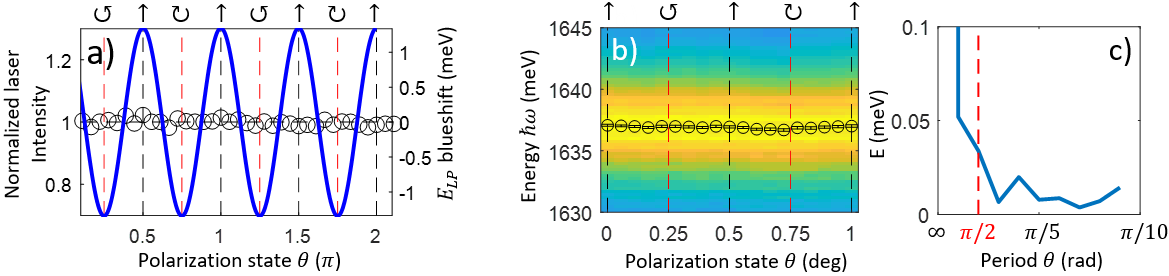}
\caption{\textbf{Polarization dependence measurements}. a) Circle symbols: laser intensity measured just before the cryostat input window versus $\theta$. solid blue line, calculated intensity modulation that would be required to achieve the spectral blueshift modulation amplitude (right axis) shown in Fig.2.d of the main text. b) Lower polariton normalized transmission spectrum versus $\theta$ (horizontal axis), in the linear response regime $W=0.11\,$pJ. The circle symbols are the extracted peak centers $E_{lp}(\theta)$. Fourier analysis of $E_{lp}(\theta)$. The dashed line show the $\pi/2$ periodicity where birefringence should show up as a peak}
\label{figS3}
\end{figure*}

\subsection{Quantifying unwanted laser intensity modulation versus~$\theta$}
Two silver mirrors (not shown in Fig.\ref{figS1}) and a doublet lens are present between the quarterwave plate and the microcavity. Their transmission and reflectivity are in principle polarization independent, but we checked them experimentally in order to prevent any bias in interpreting Fig.2.d of the main text. We thus recorded the laser intensity $I_{\rm las}(\theta)$, where $I_{\rm las}$ is measured with the thermal-head power-meter placed just before the cryostat window. The result is shown in Fig\ref{figS3}.a (circle symbols). The standard deviation is $\sigma_{I,\theta}=1\%$.

This measurement is compared with the theoretical intensity oscillation that would be required to explain the results in Fig.2.d of the main text. The blue line is the fitted spectral shift modulation shown in Fig.2.d in the main text, plotted using the right axis. It is a function of the form $B_0+B_\theta\sin ^2 2\theta$, where $B_0=7.9\,$meV and $B_\theta=-2.7\,$meV. The relation between the right (polariton energy) and left (laser intensity) vertical axes is calculated using the measured derivative of the polariton energy as a function of $W$, $\partial E_{\rm lp}/\partial W$, around $W=451\,$pJ (extracted from the dataset shown in Fig.2.a of the main text). This analysis shows that to explain the measurement in Fig.2.d with intensity modulation of the laser, a $30\%$ intensity modulation amplitude would be required (solid line in Fig.\ref{figS3}.a). Upon comparing with $\sigma_{I,\theta}$, we see that this explanation can be safely ruled out.

\subsection{Quantifying the microcavity birefringence}
We check here that the microcavity does not present any residual birefringence by measuring the lower polariton transmission spectrum $T/T_m$ at $T=105\,$K versus $\theta$, in the linear response regime ($W=0.1\,$pJ). A birefringence would show up as an oscillation of the spectrum between a single and a double peak structure versus $\theta$, of period $\pi/2$. The measurement is shown in Fig.\ref{figS3}.b, where the plot is centred on the lower polariton resonance. The peak is fitted with a Lorentzian shape, and its center $E_{\rm lp}$ is plotted versus $\theta$ as the circle symbols.

In order to be quantitative, we Fourier transform $E_{\rm lp}(\theta)$, and look at the amplitude of the $\pi/2$ periodicity (Fig.\ref{figS3}.c). We find no modulation peak at this periodicity, and the corresponding Fourier component exhibits a $50\,\mu$eV amplitude, stemming from the experimental noise. This value thus constitutes an upper bound to any polariton polarization splitting, which is thus much smaller than any linewidth.

\section{Bare cavity nonlinearity measurement}
\begin{figure}[hbt]
\includegraphics[width=\columnwidth]{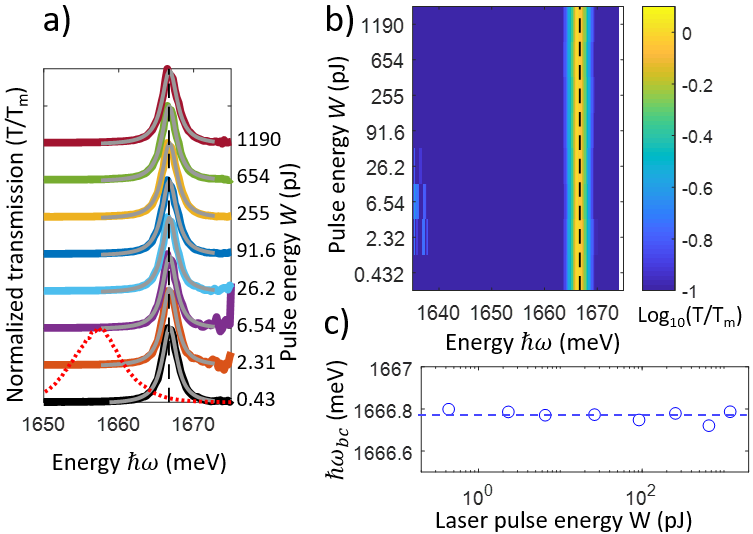}
\caption{\textbf{Spectral shift of the bare cavity resonance}. (a) Measured normalized transmission spectra $\mathcal{T(\omega)}/\mathcal{T}_m$ of the bare cavity resonance at T=105K in the area labelled (3). The spectra are stacked from the lowest used pulse energy $W$ at the bottom (black line) to the highest at the top. the pulse energy $W$ (expressed in picoJoules) used for each spectrum is indicated on the right axis. The laser pulse spectrum is shown as a red dotted line. The dashed highlight the cavity resonance peak energy. Lorentzian fits are shown as a solid gray line on each spectra. b) Same dataset as in a) shown in color log-scale. c) cavity resonance peak energy $\hbar\omega_{bc}$ versus $W$ (horizontal log scale for $W$).}
\label{figS4}
\end{figure}

We measured the nonlinearity of the bare cavity in an area of the microcavity structure where no MoSe$_2$ material, monolayer or bilayer, is present. In this area labelled (3), the cavity spacer consists only of PMMA. Fig.\ref{figS4}.a-b show a measurement of the transmission spectra in area (3) versus $W$. Each spectra is fitted with a Lorentzian, and the center energy of each peak $\hbar\omega_{bc}$ is plotted in Fig.\ref{figS4}.c versus the laser pulse energy $W$. The resulting dataset $\hbar\omega_{bc}(W)$ is constant throughout the whole range of $W$s and its standard deviation amounts to $\sigma=25\,\mu$eV. We can thus safely rule out any participation of the bare cavity material to the observed optical nonlinearites.

\section{Excitonic state effective mass and binding energy}
The excitonic effective masses, Bohr radius and binding energies in the different bulk materials mentioned in Fig.1.c-d of the main text, are found in the literature. These data are summarized in the table (\ref{table_X}).

\begin{table}[h!]
    \centering
    \begin{tabular}{ c|c|c|c|c|c|c} \hline
              &$m_e^*$& $m_{hh}^*$&$\mu_X$&$E_b$&$a_B$&Ref.\\
              &($m_e$)&($m_e$)&($m_e$)&(meV)&(nm)& \\
              \hline
         GaAs & 0.063 & 0.51 & 0.056 & 4.2 & 14.5 & \cite{GaAs_mu,GaAs_Eb,GaAs_ab} \\
         CdTe & 0.1 & 0.55 & 0.085 & 10 & 7.3 & \cite{CdTe_mu,CdTe_Eb,CdTe_ab} \\
         ZnSe & 0.21 & 0.6 & 0.15 & 20 & 4.1 & \cite{ZnSe_mu,ZnSe_Eb,ZnSe_ab} \\
         ZnO & - & - & - & -  &  1.8 & \cite{ZnO_ab} \\
         CuBr & 0.22 & 1.11 & 0.18 & 100 & 1.25 & \cite{CuBr_mu,CuBr_Eb,CuBr_ab}\\
         CuCl & 0.43 & 1.85 & 0.35 & 190 & 0.7 &\cite{CuCl_mu,CuCl_Eb,CuCl_ab}\\
         MoSe$_2$ & 0.8 & 0.5 & 0.3 & 550 & 1.1 & \cite{MoSe2_mu,MoSe2_Eb,mose2_ab}\\ \hline
    \end{tabular}
    \caption{Material parameters mentioned in this work: electron ($m_e^*$) and heavy-hole ($m_{hh}^*$) effective mass, excitonic reduced mass $\mu_X$, and binding energies $E_b$. Bulk material is considered for GaAs, CdTe, ZnSe, CuBr, and CuCl. MoSe$_2$ cited effective masses are for a single monolayer. The cited binding energy is for a monolayer grown surrounded by a graphene bilayer on one side and vacuum on the other.}
    \label{table_X}
\end{table}

 We consider that an excitonic state is stable at room temperature when its binding energy $E_b=0.5k_b\times300\,$K where the factor $0.5$ accounts for the fact that in a quantum well of thickness comparable with the Bohr radius, $E_b$ is typically twice that in the bulk \cite{Mathieu:1992}.

\subsection{A note on the interaction constants in the hydrogenic exciton theory}

In the HE picture, the excitonic Bohr radius depends on the reduced mass $\mu$ and dielectric constant $\epsilon_r$ as $a_B=\epsilon_r/\mu\times 2\pi\epsilon_0\hbar^2/e^2$, where $\epsilon_0$ is the vacuum permittivity, and $e$ is the electron charge. $g_s$ can thus be rewritten as
\begin{equation}
g_s=C_0\frac{\Omega\epsilon_r^2}{\mu^2},
\label{eq_gs1}
\end{equation}
where $C_0=\epsilon_0^216\pi^2\hbar^4/(7e^4)$. Using $g_x=6\hbar^2/(2\mu)$, $g_s$ is thus connected to $g_x$ as
\begin{equation}
g_s=C_1\Omega\epsilon_r^2g_x^2,
\label{eq_gs2}
\end{equation}
where $C_1=\epsilon_0^216\pi^2/(63e^4)$. We can draw two considerations from there:
\begin{itemize}
 \item While $g_x$ depends only on the exciton reduced mass $\mu$, Eq.(\ref{eq_gs1}) shows that $g_s$ depends on the dielectric constant as $\epsilon_r^2$. $g_s$ is thus much more sensitive than $g_x$ to any change in $\epsilon_r$.
 \item As is shown by Eq.(\ref{eq_gs2}), a correction of $g_x$ by a factor $\alpha_0$ thus results in a correction of $g_s$ by a factor $\alpha_0^2$.
\end{itemize}

\section{Equations of motion for exciton and photon fields in the $\theta$-polarization basis}
The dynamics of the system can be described by the strongly-coupled Exciton-Photon Hamiltonian, including the spin anisotropic two-body contact interactions between excitons and the saturation interaction due to the finite exciton oscillator strength \cite{Rochat:2000,Glazov:2009}.
\begin{equation}
\hat{\cal H}= \hat {\cal H}_{\sf{lin}} + \hat {\cal H}_{\sf{int}}
\end{equation}
Here $ \hat {\cal H}_{\sf{lin}}$ is the linear Hopfield part and $ \hat {\cal H}_{\sf{int}}$ is the non-linear interaction part which are given as
\begin{eqnarray*}
\hat {\cal H}_{\sf{lin}} &=& \int d\mathbf{r}\, \sum_{\sigma} \Big[ \hat{\psi}_{\rm x,\sigma}^{\dagger}(\mathbf{r})\, \left( \hbar \omega_{\rm x} \right)  \, \hat{\psi}_{\rm x,\sigma}(\mathbf{r})  \\
&& \, +  \,\hat{\psi}_{\rm c,\sigma}^{\dagger}(\mathbf{r})\left( \hbar \omega_{\rm c, 0} -\frac {\hbar ^{2}\nabla^{2}} {2m_{\rm c}}\right) \hat{\psi}_{\rm c,\sigma}(\mathbf{r}) \\
&& \,  + \,\frac {\hbar \Omega } {2}\left( \hat{\psi}_{\rm x,\sigma}^{\dagger}(\mathbf{r})\,\hat{\psi}_{\rm c,\sigma}(\mathbf{r}) + \mathbf{h.c.}\right) \Big]\\[0.2cm]
\hat {\cal H}_{\sf{int}} &=& \int d\mathbf{r}\, \sum_{\sigma} \Bigg[ \left(\frac{\hbar\,g_{x,\parallel}}{2}\right)\hat{\psi}_{\rm x,\sigma}^{\dagger}(\mathbf{r})\,\hat{\psi}_{\rm x,\sigma}^{\dagger}(\mathbf{r})\, \hat{\psi}_{\rm x,\sigma}(\mathbf{r})\, \hat{\psi}_{\rm x,\sigma}(\mathbf{r}) \\
&& \, - \,\left(\frac{\hbar g_{s}}{2}\right) \left( \hat{\psi}_{\rm c,\sigma}^{\dagger}(\mathbf{r})\,\hat{\psi}_{\rm x,\sigma}^{\dagger}(\mathbf{r})\, \hat{\psi}_{\rm x,\sigma}(\mathbf{r})\, \hat{\psi}_{\rm x,\sigma}(\mathbf{r}) + \mathbf{h.c.}\right)\Bigg]\\
&& \,  + \, \hbar g_{x,\perp} \,\hat{\psi}_{\rm x,+}^{\dagger}(\mathbf{r})\,\hat{\psi}_{\rm x,-}^{\dagger}(\mathbf{r})\, \hat{\psi}_{\rm x,-}(\mathbf{r})\, \hat{\psi}_{\rm x,+}(\mathbf{r})
\end{eqnarray*}
where $\hat \psi_{\alpha,\sigma}$ are the exciton ($\alpha=\rm{x}$) and cavity photon ($\alpha=\rm{c}$) field operators with circular polarization $\sigma=\{+,-\}$ respectively, satisfying bosonic commutation relations.   $\omega_{\rm c,0}$ is the cavity photon frequency at vanishing in-plane wavevector $k_\parallel=0$ and $m_{\rm c}$ its effective mass. $\omega_{\rm x}$ is the excitonic transition frequency, of which we neglect the kinetic contribution within the light cone. The excitonic level and cavity resonance are taken as polarization-isotropic.  $\hbar\Omega$ is the Rabi splitting,  $g_s$ is the saturation interaction constant, $g_{x,\parallel}$ and $g_{x,\perp}$ are the Coulomb interaction constants between exciton of parallel and opposite spin respectively.

We derive the Heisenberg equation of motion as $i \hbar \partial_t \hat{\psi}_{\alpha,\sigma}=[\hat{\psi}_{\alpha,\sigma},{\cal \hat{H}}],$ and take the mean field approximation
$\langle \hat \psi_{\alpha,\sigma}\rangle= \psi_{\alpha,\sigma}$. Using the input-output theory to include the pump and losses \cite{Ciuti:2006}, the equations of motion read
\begin{eqnarray}
i\partial_t \psi_{\rm c,+} &=& \left(\omega_{\rm c,0}-\frac{\hbar}{2m}\nabla^2 - i\frac{\gamma_c}{2}\right)\psi_{\rm c,+}  \nonumber \\
&& + \left(\frac{\Omega}{2}-\frac{g_s}{2}|\psi_{\rm x,+}|^2 \right)\psi_{\rm x,+}+\sqrt{2\gamma_{\rm in}}A_{\rm in,+} \nonumber \\
i\partial_t \psi_{\rm x,+} &=& \left(\omega_{\rm x} - i\frac{\gamma_x}{2}+g_{x,\parallel}|\psi_{\rm x,+}|^2+g_{x,\perp}|\psi_{\rm x,-}|^2\right)\psi_{\rm x,+}  \nonumber \\
&& + \left(\frac{\Omega}{2}-g_s|\psi_{\rm x,+}|^2 \right)\psi_{\rm c,+}-\frac{g_s}{2}\psi_{\rm x,+}^2\psi_{\rm c,+}^* \nonumber \\
i\partial_t \psi_{\rm c,-} &=& \left(\omega_{\rm c,0}-\frac{\hbar}{2m}\nabla^2 - i\frac{\gamma_c}{2}\right)\psi_{\rm c,-}  \nonumber \\
&& + \left(\frac{\Omega}{2}-\frac{g_s}{2}|\psi_{\rm x,-}|^2 \right)\psi_{\rm x,-}+\sqrt{2\gamma_{\rm in}}A_{\rm in,-} \nonumber \\
i\partial_t \psi_{\rm x,-} &=& \left(\omega_{\rm x} - i\frac{\gamma_x}{2}+g_{x,\parallel}|\psi_{\rm x,-}|^2+g_{x,\perp}|\psi_{\rm x,+}|^2\right)\psi_{\rm x,-}  \nonumber \\
&& + \left(\frac{\Omega}{2}-g_s|\psi_{\rm x,-}|^2 \right)\psi_{\rm c,-}-\frac{g_s}{2}\psi_{\rm x,-}^2\psi_{\rm c,-}^*, \nonumber
\label{eq:mf_vec}
\end{eqnarray}
where
$A_{\rm in,\pm}(\mathbf{r},t)$ describes the incident laser pulse field density, , $\gamma_c=\gamma_{\rm in}+\gamma_{\rm out}$ is the cavity radiative decay rate, and $\gamma_{\rm in}$ ($\gamma_{\rm out}$) are the coupling rate of the laser into the cavity (of the cavity into the outside detection channel).

In order to describe the experimental conditions, we rewrite these equations in the $|\theta,\bar{\theta}\rangle$ basis where, $\theta$ is the rotation angle of the quarter waveplate with respect the linear polarization $|y\rangle$ which is that of the laser output. Algebra manipulation leads to following basis transformation
\begin{align}
|\theta\rangle=\frac{1}{2}(i+e^{i2\theta})|+\rangle -\frac{1}{2}(i+e^{-i2\theta})|-\rangle \\
|\bar{\theta}\rangle=\frac{1}{2}(-i+e^{i2\theta})|+\rangle + \frac{1}{2}(-i+e^{-i2\theta})|-\rangle,
\end{align}
which allows us rewriting the equations of motion in the $(\theta,\bar{\theta})$ basis. In this transformation, we use the fact that in the experiment, the source term $A_{in,\bar{\theta}}$=0. Moreover, each nonlinear term involves products of three fields of the form $\psi_{\theta,\bar{\theta}}\psi^*_{\theta,\bar{\theta}}\psi_{\theta,\bar{\theta}}$, such that none of them can serve as auxiliary source terms for the $\bar{\theta}$ components of the field, as long as its initial amplitude is zero. As a result, only the nonlinear terms of the form $\psi_\theta\psi^*_\theta\psi_\theta$ are nonzero, and we can drop the two equations of motions describing the $\bar{\theta}$-polarized field. We thus obtain the equations of motion for the $\theta$-polarized fields as
\begin{align}
\label{eq:mf_vec1}
i\partial_t \psi_{\rm c,\theta}&= \left(\omega_{\rm c,0}-\frac{\hbar}{2m}\nabla^2 - i\frac{\gamma_c}{2}\right)\psi_{\rm c,\theta}+\sqrt{2\gamma_{\rm in}}A_{\rm in,\theta}  \nonumber \\
&+\left(\frac{\Omega}{2}-\frac{\tilde{g}_s(\theta)}{2}|\psi_{\rm x,\theta}|^2\right)\psi_{\rm x,\theta} \\
\label{eq:mf_vec2}
i\partial_t \psi_{\rm x,\theta}&= \left(\omega_{\rm x} - i\frac{\gamma_x}{2}+\tilde{g}_x(\theta)|\psi_{\rm x,\theta}|^2\right)\psi_{\rm x,\theta} \nonumber \\
&+ \left(\frac{\Omega}{2}-\tilde{g}_s(\theta)|\psi_{\rm x,\theta}|^2 \right)\psi_{\rm c,\theta}-\frac{\tilde{g}_s(\theta)}{2}\psi_{x,\theta}^2\psi_{c,\theta}^*
\end{align}
where
\begin{align}
\tilde{g}_x(\theta)&=\frac{g_{x,\parallel}+g_{x,\perp}}{2}+\frac{g_{x,\parallel}-g_{x,\perp}}{2}\sin^2(2\theta) \\
\tilde{g_s}(\theta)&=\frac{g_s}{2}\left[(1+\sin^2(2\theta)\right],
\end{align}
corresponding to Eqs.(1-4) of the main text.

\subsection{Polarization dependent spectral shift}

\subsubsection{Equations of motion in the polariton basis}
In order to better understand the $\theta$-dependence of the spectral shift that we observe in Fig.2.d in the main text, we can further transform the equations of motion to write them into the polariton basis.
\begin{align}
|L\rangle&=X|x\rangle -C|c\rangle \\
|U\rangle&=C|x\rangle + X|c\rangle
\end{align}
is the polaritons basis expressed in the exciton photon basis. $|L,U\rangle$ are the upper and lower polariton states, $|x,c\rangle$ are the exciton and cavity photon states, and $(X,C)$ are the usual excitonic and photonic Hopfield coefficients. The equation of motions in this basis read
\begin{align}
\label{eq:mf_vec1}
i\partial_t \psi_{\rm L,\theta}&= \left(\omega_{\rm L,0}-\frac{\hbar}{2m_L}\nabla^2 - i\frac{\gamma_L}{2}\right)\psi_{\rm L,\theta}  \nonumber \\
&+\sqrt{2\gamma_{\rm in}}CA_{\rm L,in}-C A_{L,in}+\Delta_L \\
\label{eq:mf_vec2}
i\partial_t \psi_{\rm U,\theta}&= \left(\omega_{\rm U,0}-\frac{\hbar}{2m_U}\nabla^2 - i\frac{\gamma_U}{2}\right)\psi_{\rm U,\theta} \nonumber \\
&+\sqrt{2\gamma_{\rm in}}XA_{\rm U,in}+XA_{U,in}+\Delta_U,
\end{align}
where $\hbar\omega_{L,U,0}$, $\hbar\gamma_{L,U}$, and $m_{L,U}$ are the lower and upper polaritons rest energy, linewidth and effective mass respectively. $A_{L,U,in}$ is the laser amplitude at the lower and upper polaritons energy. $\Delta_{U/L}$ gather all the nonlinear terms in the upper (U) and lower (L) polariton field equations.
 After discarding  the terms that average to zero, $\Delta_{U/L}$ read
 \begin{widetext}
\begin{align}
& \Delta_L = \Bigg[\frac{1}{2}\left(X^2|\psi_L|^2 + 2C^2|\psi_U|^2\right)\left[\left(X^2g_{x,\parallel}+X^2g_{x,\perp}+2XCg_s\right)+\left(X^2g_{x,\parallel}-X^2g_{x,\perp}+2XCg_s\right)\sin^2 2\theta \right] \nonumber \\ \label{LPBS_v_theta}
& - g_sXC|\psi_U|^2(1+\sin^2 2\theta) \Bigg]\psi_L \\
& \Delta_U = \Bigg[\frac{1}{2}\left(C^2|\psi_U|^2+2X^2|\psi_L|^2\right)\left[\left(C^2g_{x,\parallel}+C^2g_{x,\perp}-2XCg_s\right)+\left(C^2g_{x,\parallel}-C^2g_{x,\perp}-2XCg_s\right)\sin^2 2\theta\right] \nonumber \\
& + g_sXC|\psi_L|^2(1+\sin^2 2\theta)\Bigg]\psi_U
\end{align}
\end{widetext}
These equation allow a very general understanding of the different contributions to the spectral shifts of the lower and upper polaritons, including when a comparable population occupies the lower and upper polariton states, like in our case.